\documentstyle[prl,aps,graphicx]{revtex}

\begin{document}

\title{Momentum-Transfer to and Elementary-Excitations of a
Bose-Einstein Condensate by a Time-Dependent Optical Potential}

\author{Y.\ B.\ Band and M.\ Sokuler}
\address{
Departments of Chemistry and Physics, Ben-Gurion
University of the Negev, Beer-Sheva, Israel  \, 84105
}

\maketitle

\begin{abstract}
We present results of calculations on Bose-Einstein condensed
$^{87}$Rb atoms subjected to a moving standing-wave light-potential of
the form $V_L(z,t) = V_0(t) \cos(q z-\omega t)$.  We calculate the
mean-field dynamics (the order paramter) of the condensate and
determine the resulting condensate momentum in the $z$ direction,
$P_z(q,\omega,V_0,t_p)$, where $V_0$ is the peak optical potential
strength and $t_p$ is the pulse duration.  Although the local density
approximation for the Bogoliubov excitation spectral distribution is a
good approximation for very low optical intensities, long pulse
duration and sufficiently large values of the wavevector $q$ of the
light-potential, for small $q$, short duration pulses, or for
not-so-low intensities, the local density perturbative description of
the excitation spectrum breaks down badly, as shown by our results.
\end{abstract}

\pacs{PACS Numbers: 03.75.Fi, 03.75.-b, 67.90.+z, 71.35.Lk}

\section{Introduction}

Elementary excitations of a Bose-Einstein condensate (BEC) can be
explored with matter-wave interference studies using two-photon Bragg
pulse spectroscopy \cite{Andrews,Hagley,Anderson,Orzel,Cata,Stenger,%
Stamper,Vogels,Steinhauer,Nir}.  In such studies, the momentum
imparted to the BEC by a Raman scattering process can be studied as a
function of the temporal duration of the optical pulses, $t_p$, the
detuning of the Bragg pulses from atomic resonance, $\Delta$, the
intensity of the Bragg pulses, the difference of the Bragg pulse
wavevectors which is denoted by the wavevector ${\bf q}$, and the
difference of the central frequency of the two laser pulses $\omega$. 
In the linear response limit, the response of the BEC to a weak
perturbation with wavevector ${\bf q}$ and energy $\hbar \omega$ is
given in terms of the dynamic structure factor $S({\bf q},\omega)$
\cite{NozieresPines},
\begin{equation}
S({\bf q},\omega) = {\frac{1}{\cal Z}} \sum_{m,n} e^{-\beta E_{m}}
\mid \langle m | \delta \rho _{{\bf q}} | n \rangle \mid
^{2}\delta (\omega -\omega_{mn}) \,,
\label{eq:S}
\end{equation}
where $\beta=1/kT$, $\hbar \omega_{mn}= E_m - E_n$ is the difference
of two energy eigenvalues of the BEC, the density fluctuation
$\delta\rho_{{\bf q}}$ is induced by a perturbation with wavevector
${\bf q}$ and frequency $\omega$ that oscillates like $e^{i({\bf
q}\cdot {\bf r}-\omega t)}$, and ${\cal Z}$ is the usual partition
function.  The momentum imparted to the BEC by the light-potential and
its dependence on the wavevector ${\bf q}$ and frequency $\omega$ can
be directly related to the structure factor $S({\bf q},\omega)$, and
to the Bogoliubov dispersion relation, $E_B({\bf q})$ versus ${\bf q}$
\cite{Bogol-Beliaev}.  The excitation modes of a BEC have been
measured \cite{Jin,Mewes} and can be directly related to $S({\bf
q},\omega)$ and $E_B({\bf q})$.  The momentum transferred to a BEC by
a moving standing-wave light-potential has also been directly measured
over a wide range of ${\bf q}$ and $\omega$
\cite{Stamper,Steinhauer,Nir}, and it is of interest to calculate the
momentum transfer versus ${\bf q}$ and $\omega$ so that the
calculations can be compared with experiment.

Here we report on the results of calculations of BEC excitation by
Bragg pulse spectroscopic techniques to obtain the momentum transfer
versus ${\bf q}$ and $\omega$ for weak and strong optical excitations. 
We find that, even with what is ordinarily considered weak intensity
Bragg pulses, processes that are higher than first order (linear
response) play a role in the excitation process.  We also find that a
local density Bogoliubov description is not valid for small
wavevectors $q$.  We describe the nature of the the higher order
processes and their influence on the momentum and energy of the BEC
excitation.  We show how the simple Bogoliubov picture of the
excitation is modified over a range of momentum-transfers and
excitation energies due to higher order light-scattering processes,
finite BEC size and inhomogeneity effects.

We consider Bose-Einstein condensed $^{87}$Rb atoms in the $|F=2,
M_F=2\rangle$ hyperfine state confined in a harmonic oscillator
potential, an array of optical traps and a gravitational field, and
use parameters similar to those used in experiments carried out at the
Weizmann Institute \cite{Nir}.  The initial BEC is cigar shaped with
$N = 10^5$ atoms in a static harmonic trap potential $V_{ho}({\bf r})
= \frac{m\omega_z^2}{2} z^2 + \frac{m\omega_{x,y}^2}{2} (x^2+y^2)$
with frequencies $\omega_z = 2\pi \, \times \, 25$ Hz, $\omega_{x,y}
\equiv \omega_x = \omega_y = 2\pi \, \times \, 220$ Hz ($\bar{\omega}
\equiv (\omega_x\omega_y\omega_z)^{1/3} = 2\pi \, \times \, 106$ Hz). 
The Bragg pulses propagate with wavevectors ${\bf k}_1$ and ${\bf
k}_2$ in the $x$-$z$ plane with angles $\pm\theta/2$ relative to the
$x$ axis.  The central frequency of one pulse is greater than that of
the other, $\omega_2 = \omega_1 - \omega$, and the difference
frequency $\omega$ is controlled using an acousto-optic modulator. 
The electric field takes the form ${\bf E}(t) = {\bf E}_1(t)
\exp[i({\bf k}_1 \cdot {\bf x} -\omega_1 t) + {\bf E}_2(t) \exp[i({\bf
k}_2 \cdot {\bf x} -\omega_2 t)$, with lin $\parallel$ lin
configuration for the field polarizations and equal intensities for
the two Bragg pulses.  When the Bragg pulses are switched on, the
atoms are trapped at the antinodes of a vertically oriented,
red-detuned optical moving standing-wave; the antinodes are separated
by $\Delta z = \lambda/(2\sin(\theta/2))$.  The momentum transferred
to an atom upon absorbing a photon from the field with wavevector
${\bf k}_1$ and emitting a photon with wavevector ${\bf k}_2$ is given
by $\hbar{\bf q} = \hbar({\bf k}_1 - {\bf k}_2) = \hbar q \hat{z}$,
where the $\hbar q = 2 \hbar k_{ph} \sin(\theta/2)$, and $\hbar k_{ph}
= 2\pi\hbar/\lambda$ is the photon momentum.  Here $\lambda = 780$ nm
is the central wavelength of the Bragg pulses.  The light-potential
experienced by the atoms in the BEC as a result of the Bragg pulses is
given by $V_L(z,t) = V_0(t) \cos(q z-\omega t)$.  The well-depth of
the optical potential, $V_0(t)$, is proportional to the intensity of
the Bragg pulses and inversely proportional to the detuning from
resonance, $\Delta$, i.e., $V_0(t) = \frac{\hbar \Omega_1(t)
\Omega_2(t)}{4\Delta}$ where $\Omega_1(t)$ and $\Omega_2(t)$ are the
Rabi frequencies.  The well-depth temporal dependence $f(t)$, where
$V_0(t) = V_0 \, f(t)$, is taken to have a Gaussian rise-time and
fall-time of width $t_r = 20 \, \mu$s; $V_0(t)$ is constant for a time
duration $t_p$ between the rise and fall ($f(t) = 1$ for the time
interval $t_p$ so $V_0(t) = V_0$ in this interval).  Pulses with short
($t_p = 1$ ms) and long pulse duration ($t_p = 6$ ms) are used.  The
strength of $V_0$ that is used in the calculations will be specified
in recoil units, $E_R \equiv (\hbar k_{ph})^2/(2m)$.  Absorption of a
photon from one pulse and stimulated emission into the other pulse
produces a perturbation with energy $\hbar \omega$ and momentum $\hbar
q \hat{z}$.  In the experiments, the light-potential and the harmonic
potential are both switched-off (dropped), releasing the atoms to fall
under the influence of gravity, and absorption images are taken after
the particles evolve for some specified period of time under the
influence of gravity.

\section{Theoretical Formulation}

We calculate the dynamics within a mean-field treatment using the
time-dependent Gross-Pitaevskii equation (GPE)
\cite{Gross,Pitaevskii},
\begin{equation}
i\hbar \frac{\partial\psi({\bf r},t)}{\partial t} = (\frac{p^2}{2m} +
V({\bf r},t) + g|\psi|^2) \psi \,, \label{GPE}
\end{equation}
where
\begin{equation}
V({\bf r},t) = V_{ho}({\bf r}) - mgz + V_L(z,t) \,,  \label{Poten}
\end{equation}
is the potential, and $g = \frac{4\pi N a_{0}\hbar^{2}}{m}$ is the
atom-atom nonlinear interaction strength which is proportional to the
$s$-wave scattering length $a_{0}$ and the total number of condensed
atoms $N$.  The wave function (order parameter) $\psi({\bf r},t)$ is
propagated with a split-operator fast Fourier transform method.  Due
to the large number of grid points necessary in the lattice direction
($z$), we found it necessary to convert the 3D GPE into an effective
1D GPE with similar dynamics as described in Ref.~\cite{BandTripp}. 
The wave function in momentum space, $\psi(k_z,t)$, is determined by
taking the Fourier transform of $\psi(z,t)$.  The net momentum of the
BEC at any time $t$ is given by
\begin{equation}
P_z(t) = \int_{-\infty}^{\infty} dk_z \psi^{*}(k_z,t) (\hbar k_z) 
\psi(k_z,t) \,. \label{Pz}
\end{equation}
The expectation values of all dynamical quantities (e.g., energies,
$\langle\Delta z\rangle$, etc.)  can be easily determined using the
calculated wavepackets in either position or momentum space.

A perturbative estimate of the rate of momentum-transfer to the BEC by
the Bragg pulses is given by \cite{Brunello}
\begin{equation}
{\frac{dP_z(t)}{dt}} = \frac{2\pi}{\hbar} q
\left(\frac{V_0}{2}\right)^2 \left[ S({\bf q},\omega)-S(-{\bf
q},-\omega )\right] \,.  \label{dPdt}
\end{equation}
For a uniform zero-temperature BEC, the Bogoliubov excitation energy
at momentum-transfer ${\bf q}$ is
\begin{equation}
E_B(q) \equiv \hbar \Omega_B(q) = \sqrt{\epsilon(q) [\epsilon(q) + 2 g
n]}\,,
\end{equation}
where $\epsilon(q) = (\hbar q)^2/2m$, $n$ is the density, and the
dynamic structure factor is given by $S({\bf q},\omega) =
\frac{\epsilon(q)}{E_B(q)} \delta(\hbar\omega - E_B(q))$.  Hence, the 
momentum transferred is given by
\begin{equation}
P_z(t) = \frac{\pi}{2\hbar} \left( \int_0^t \,dt^{\prime} 
V_0^2(t^{\prime}) \right) 
\frac{q\epsilon(q)}{\sqrt{\epsilon(q) [\epsilon(q) + 2 g n]}} \,
\delta(\hbar\omega - E_B(q))\,.  \label{P_z(t)}
\end{equation}
Thus, the momentum transferred by a moving standing-wave excitation to
a uniform condensate is non-vanishing only when $\omega = \Omega_B(q)$,
and its magnitude is proportional to the product of the integral over
time of $V_0^2$ and $\frac{q\epsilon(q)}{E_B(q)}$.

For a non-uniform BEC with density profile that varies smoothly with
position, one can define a local Bogoliubov excitation energy,
\begin{equation}
E_B(q,{\bf r}) \equiv \hbar \Omega_B(q,{\bf r}) = \sqrt{\epsilon(q)
[\epsilon(q) + 2 g n({\bf r})]} \,,
\end{equation}
and a local density approximation dynamic structure factor
that behaves locally as a uniform gas \cite{Stamper}:
\begin{equation}
S_{LDA}({\bf q},\omega) = N^{-1} \int d{\bf r} \, n({\bf r})
\frac{\epsilon(q)} {E_B(q,{\bf r})} \, \delta(\hbar\omega - E_B(q,{\bf
r})) \,.
\label{Sqomegau}
\end{equation}
Hence, in perturbation theory (i.e., small $V_0$), the momentum
transferred to the non-uniform condensate can be approximated by
substituting $S_{LDA}({\bf q},\omega)$ for $S({\bf q},\omega)$ on the
right hand side of Eq.~(\ref{dPdt}).  Thus, momentum transfer of a
non-uniform condensate via a moving standing-wave excitation is
smeared over a range of frequencies around $\omega = \Omega_B(q)$, and
its magnitude is determined by carrying out an average of
$\frac{q\epsilon(q)}{E_B(q,{\bf r})}$ over the local density.

\section{Results}

Calculations were carried out over a range of momentum-transfer $q$
(i.e., a range of angles $\theta$), frequencies $\omega$, pulse
duration times $t_p$, and laser pulse intensities (potential strengths
$V_0$).

\begin{figure}[!htb]
%\centerline{\includegraphics[width=3in,angle=270,keepaspectratio]{N_D225_f1.eps}}
\centerline{\includegraphics[width=3in,keepaspectratio]{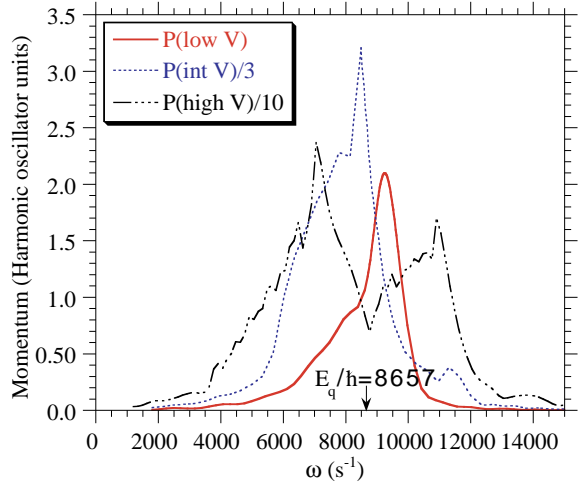}}
\caption {Momentum $P_z(q,\omega,V_0,t_p)$ versus $\omega$ for $\theta =
22.5^{\text{o}}$ ($q = 0.390 \hbar k_{ph}$) and pulse duration $t_p =
6$ ms.  Results for three different laser intensities corresponding to
potential well depths $V_0 = 0.0054$ (low), 0.0162 (intermediate) and 
0.054 (high) $E_R$ are shown.}
\label{fig1}
\end{figure}

\begin{figure}[!htb]
\centerline{\includegraphics[width=3in,angle=270,keepaspectratio]{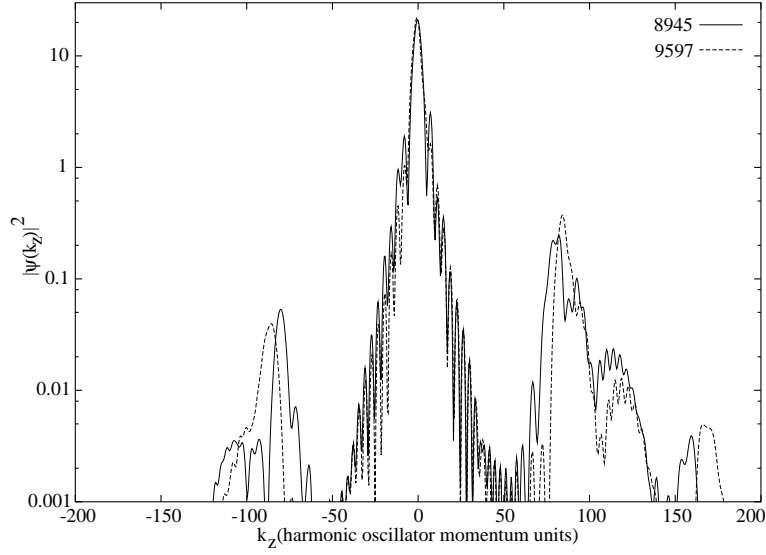}}
\caption {$|\psi(k_z,t)|^2$ vs.~$k_z$ for the low intensity case
appearing in Fig.~\ref{fig1} at $\omega = 8945$ and 9597
s$^{-1}$.  The units of $k_z$ are in harmonic oscillator momentum
units $\sqrt{\hbar m \omega_z} (= 2.44 \times 10^{-3} \, \hbar
k_{ph})$.}
\label{fig2}
\end{figure}

Fig.~\ref{fig1} shows the calculated momentum imparted to the BEC by
the Bragg pulses, $P_z(q,\omega,V_0,t_p)$, as a function of $\omega$
for $\theta = 22.5^{\text{o}}$ ($q = 0.390 \, \hbar k_{ph} = 80
\sqrt{\hbar m \omega_z}$) and pulse duration $t_p = 6$ ms.  The
calculations were performed for three different laser intensities
corresponding to potential well depths $V_0 = 0.0054$, 0.0162 and
0.054 $E_R$.  The low intensity curve peaks near the Bogoliubov
frequency $\Omega_B(q) = 8657$ s$^{-1}$.  This peak has a tail at
lower $\omega$ which is due to the non-uniform density of the BEC;
$E_B(q,{\bf r}$) for positions away from the center of the BEC is
smaller than at the center and this can give rise to the tail, as can
be understood from Eqs.~(\ref{Sqomegau}) and (\ref{dPdt}).  At
intermediate and high $V_0$, a power broadening of the spectral
distribution $P_z(\omega)$ is evident in Fig.~\ref{fig1}; at these
values of $V_0$, higher order (nonlinear) processes that populate $\pm
2 q$ momenta take place (see Fig.~\ref{fig2}).  These phenomena can
not be understood from a perturbative treatment.  At even higher
values of $V_0$, the spectral distribution becomes even wider and the
peak structure becomes even more complicated and ragged.

Fig.~\ref{fig2} shows $|\psi(k_z,t)|^2$ vs.~$k_z$ for the low
intensity case appearing in Fig.~\ref{fig1} for $\omega = 8945$
and 9597 s$^{-1}$.  The net positive momentum, ($P_z(\omega) > 0$)
resulting at these frequencies in Fig.~\ref{fig1} is due to the fact
that the peak at $k_z = q \, (\approx 0.390 \, \hbar k_{ph} = 80
\sqrt{\hbar m \omega_z}$) is larger than that at $k_z = -q$.  The peak
near $k_z = q$ has an additional high-frequency feature at around $k_z
= 120 \sqrt{\hbar m \omega_z}$; the origin of this feature is not
clearly understood.  The peaks at $k_z = \pm 2q$ are almost two orders
of magnitude reduced compared with the $k_z = \pm q$ peaks.  As the
intensity of the lasers (the potential well depths) increase, the size
of the $k_z = \pm 2q$ peaks grow in comparison with the $k_z = \pm q$
peaks, and $|\psi(k_z,t)|^2$ grows at intermediate values of $k_z$
between the peaks at $k_z = \pm jq$, where $j$ is a positive integer.

\begin{figure}[!htb]
\centerline{\includegraphics[width=3in,angle=270,keepaspectratio]{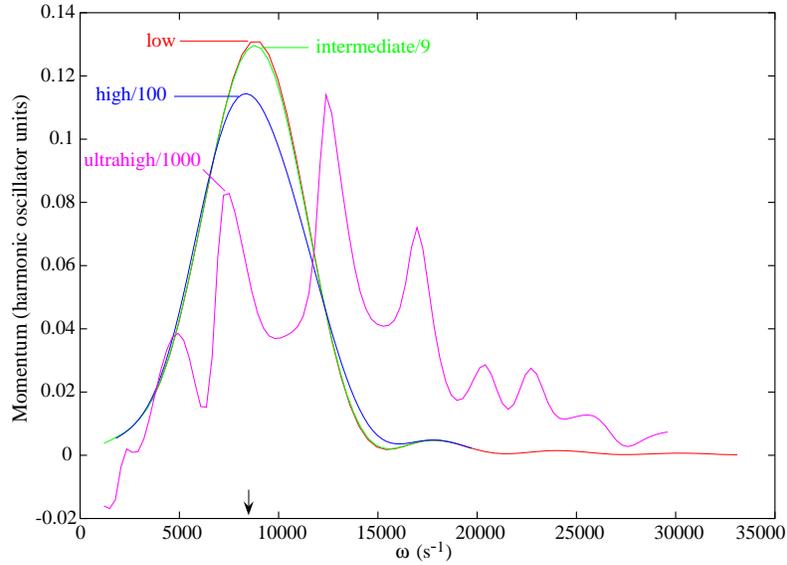}}
\caption {Same as Fig.~\ref{fig1}, except the pulse duration is 1 ms
and an additional well depth of $V_0 = 0.54$ (ultrahigh) $E_R$ is
included.  The arrow indicates the position of $\Omega_B(q)$.}
\label{fig3}
\end{figure}

Fig.~\ref{fig3} is similar to Fig.~\ref{fig1}, except the pulse
duration was taken to be 1 ms.  The width of the distribution
$P_z(\omega)$ as a function of $\omega$ for $t_p = 1$ ms is
considerably wider than for $t_p = 6$ ms for the low and intermediate
intensities.  The shorter temporal duration and therefore larger
bandwidth of the 1 ms Bragg pulses allows for a wider distribution in
frequency of $P_z(\omega)$ versus $\omega$.  The width hides the tail
of the distribution at lower $\omega$ due to the non-uniform BEC
density.  For the high intensity case, the distribution does not
change very much from the low and intermediate cases.  Moreover, it is
narrower than the high intensity 6 ms result shown in Fig.~\ref{fig1}
because the pulse fluence is smaller, hence higher order processes do
not significantly broaden the distribution.  However, for the
ultrahigh intensity case, power broadening of the distribution is
significant.  Note that the perturbation theory expression
(\ref{dPdt}) can not account for the time-domain broadening shown in
Fig.~\ref{fig3} (since it is derived assuming that the spectral width
of $V_0$ is within that of $S({\bf q},\omega)$).  Eq.~(\ref{dPdt})
shoud be modified to account for the bandwidth of the optical pulse:
\begin{equation}
{\frac{dP_z(t)}{dt}} = \frac{\pi}{2\hbar} q \left( \frac{1}{\sqrt{2\pi}} \int
d\omega \, e^{-i\omega t} |V_0(\omega)|^2 S({\bf q},\omega) + c.c. 
\right) \,. 
\label{dPdtn}
\end{equation}
This equation should then be integrated over time to obtain an 
expression that replaces (\ref{P_z(t)}) for $P_z(t)$.

\begin{figure}[!htb]
\centerline{\includegraphics[width=3in,angle=270,keepaspectratio]{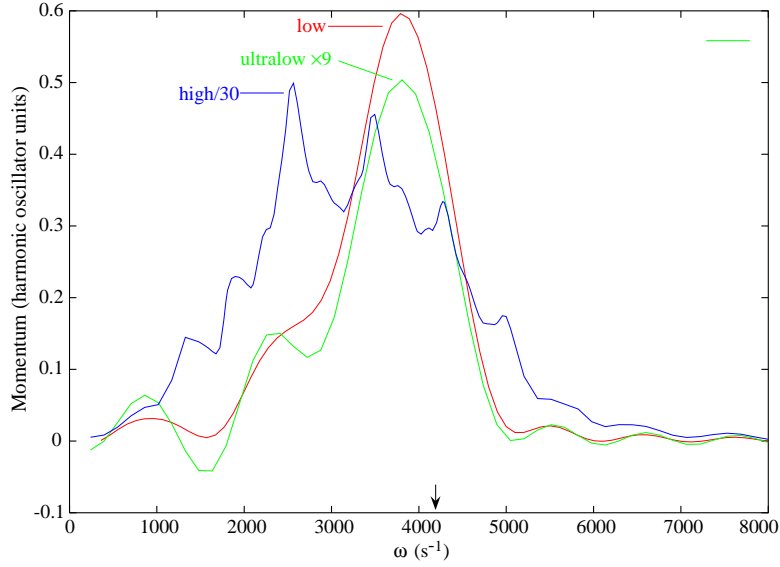}}
\caption {Same as Fig.~\ref{fig1} except for 10$^{\text{o}}$ ($q =
0.174 \, \hbar k_{ph}$).  Results for three laser intensities
corresponding to potential well depths $V_0 = 0.00162$, 0.0054 and
0.054 $E_R$ are shown.  The arrow indicates the position of $\Omega_B(q)$.}
\label{fig4}
\end{figure}

\begin{figure}[!htb]
\centerline{\includegraphics[width=3in,angle=270,keepaspectratio]{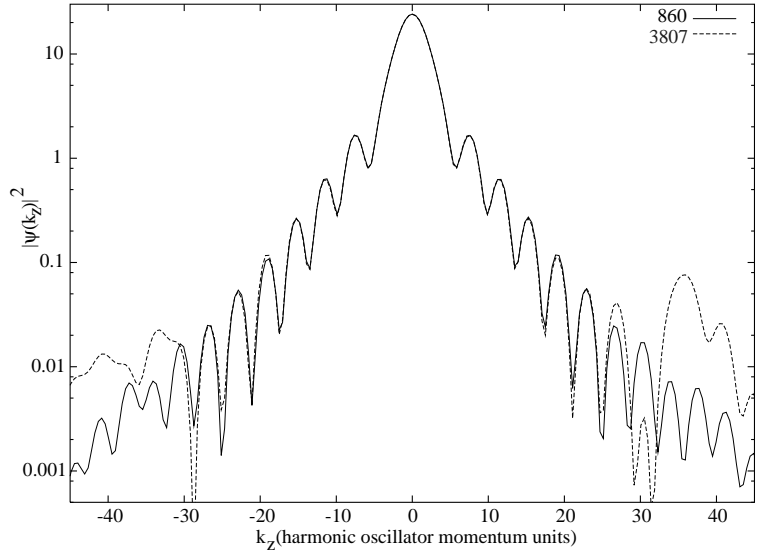}}
\caption {$|\psi(k_z,t)|^2$ vs.~$k_z$, for the ultralow intensity case
shown in Fig.~\ref{fig4} at $\omega = 860$ and 3807 s$^{-1}$.  The
units of $k_z$ are in harmonic oscillator momentum units $\sqrt{\hbar
m \omega_z} (= 2.44 \times 10^{-3} \, \hbar k_{ph})$.}
\label{fig5}
\end{figure}

We now consider smaller values of momentum-transfer $q$ such that the
excitation wavelengths of the optical potential are comparable or
larger than the size of the condensate.  Fig.~\ref{fig4} shows $P_z$
versus $\omega$ for $\theta = 10^{\text{o}}$ ($q = 0.174 \, \hbar
k_{ph} = 35.8 \sqrt{\hbar m \omega_z}$) and $t_p = 6$ ms.  The curve
for low intensity, $V_0 = 0.0054 \, E_R$, is similar to the low
intensity result in Fig.~\ref{fig1} in the sense that a peak exists
near the Bogoliubov frequency $\Omega_B(q) = 4164$ s$^{-1}$ and a low
frequency tail is present.  Increasing the intensity to $V_0 = 0.054
\, E_R$ (the curved labeled high), yields a multi-peaked ragged
spectrum due to higher order nonlinear processes.  Upon reducing the
optical potential by a factor of 0.3 from the low intensity case to
$V_0 = 0.00162 \, E_R$ (the curve labeled ultralow), new features in
the spectrum become clear.  The small feature near $\omega = 860$
s$^{-1}$ and the tail of the Bogoliubov peak at 2411 s$^{-1}$, which
was presumably due to an inhomogeneous density effect, become two
ancillary peaks.  The reason for the structure in the spectrum for the
ultralow (and low) intensity can be understood by looking at
$|\psi(k_z,t)|^2$ vs.~$k_z$ for this case as shown in Fig.~\ref{fig5}. 
Peaks at $k_z = q = 35.8 \sqrt{\hbar m \omega_z}$ are clearly seen for
$\omega = 3807$ s$^{-1}$ and these peaks fall within the tail of the
oscillatory structures associated with the broadened $k_z = 0$
condensate.  These structures at $k_z = q$ are responsible for the
peaks in the spectral distribution in Fig~\ref{fig4}.  The peak near
$\omega = 860$ s$^{-1}$ and the minimum near $\omega = 1636$ s$^{-1}$
in Fig.~\ref{fig4} result due to subtle interference of the $k_z = q$
peak with the structure surrounding the central ($k_z = 0$) peak in
Fig.~\ref{fig5}.  This kind of interference can occur when $q$ (angle
$\theta$) is sufficiently small that the peak at $k_z = q$ is within
the structure of the central peak.  It does not occur at $\theta =
22.5^{\text{o}}$ ($q = 0.390 \, \hbar k_{ph}$), rather only for angles
$\theta \le 10^{\text{o}}$.

\begin{figure}[!htb]
\centerline{\includegraphics[width=3in,angle=270,keepaspectratio]{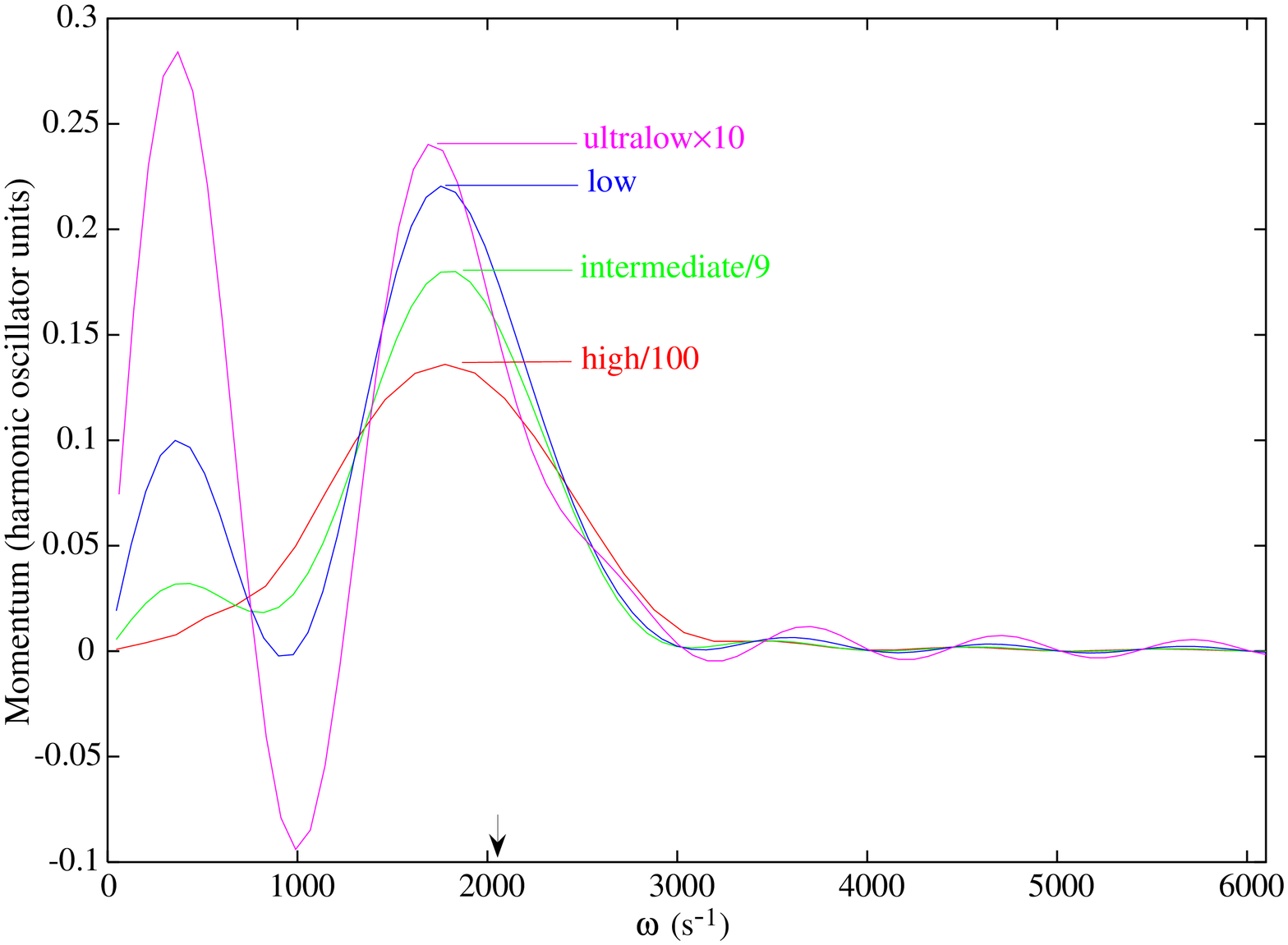}}
\caption {Same as Fig.~\ref{fig1} except for 5$^{\text{o}}$ ($q = 8.72
\times 10^{-2} \, \hbar k_{ph}$).  Results for four different laser
intensities corresponding to potential well depths $V_0 = 0.00162$,
0.0054 0.0162 and 0.054 $E_R$ are shown.  The arrow indicates position of
$\Omega_B(q)$.}
\label{fig6}
\end{figure}

In Fig.~\ref{fig6} we plot $P_z$ versus $\omega$ for $\theta =
5^{\text{o}}$ ($q = 8.72 \times 10^{-2} \, \hbar k_{ph} = 18 \,
\sqrt{\hbar m \omega_z}$), $t_p = 6$ ms and potential well depths $V_0
= 0.00162$, 0.0054, 0.0162 and 0.054 $E_R$.  At ultralow, low and
intermediate intensities, an additional peak appears at $\omega
\approx 400$ s$^{-1}$.  At all but ultralow intensity, the main peak
in the spectrum is near the Bogoliubov frequency $\Omega_B(q) = 2061$
s$^{-1}$.  At ultralow intensity the peak at $\omega \approx 400$
s$^{-1}$ is even larger than that at $\omega \approx 2000$ s$^{-1}$
and $P_z$ between these peaks becomes negative.  Again, interference
effects arising for reasons explained in connection with
Fig.~\ref{fig5} are apparently responsible.
% Fig.~\ref{fig7} shows $|\psi(k_z)|^2$ vs.~$k_z$ for three values of
% $\omega = 370$, 990 and 1687 s$^{-1}$ for the ultralow intensity case
% in Fig.~\ref{fig6}.  The peak at $k_z = q \, (= 18 \sqrt{\hbar m
% \omega_z}$) occurring within the central ($k_z = 0$) peak is somewhat
% difficult to distinguish from the fringe structures of the central
% peak for all plotted values of $\omega$.  The height of the peaks at
% $k_z = \pm q$ are not very different from one another; this accounts
% for the small values of $P_z$ obtained in Fig.~\ref{fig6}.  For the
% low intensity case (not shown in Fig.~\ref{fig7}), as opposed to the
% ultralow intensity case, the peaks in $|\psi(k_z)|^2$ at $k_z = 2q =
% 36 \sqrt{\hbar m \omega_z}$ for $\omega = 900$ and 1687 s$^{-1}$ can
% be distinguished as somewhat larger than the structure within the
% central peak; nevertheless, they are more than an order of magnitude
% lower than the peaks at $k_z = q$.
A complicated interference pattern appears around the values of $k_z =
\pm q$ in $|\psi(k_z,t)|^2$ vs.~$k_z$ (figure not shown).  This
interference plays a role in the determination of the spectrum shown
in Fig.~\ref{fig6}.  The peaks in $|\psi(k_z)|^2$ at $k_z = q$ for
$\omega = 370$ and 1687 s$^{-1}$ are larger than the peak at $k_z = q$
for $\omega = 990$ s$^{-1}$, and therefore maxima occur in the
distribution shown in Fig.~\ref{fig6} at $\omega = 370$ and 1687
s$^{-1}$ and a minimum between the peaks in the spectrum occurs for
$\omega = 990$ s$^{-1}$.

\begin{figure}[!htb]
\centerline{\includegraphics[width=3in,angle=270,keepaspectratio]{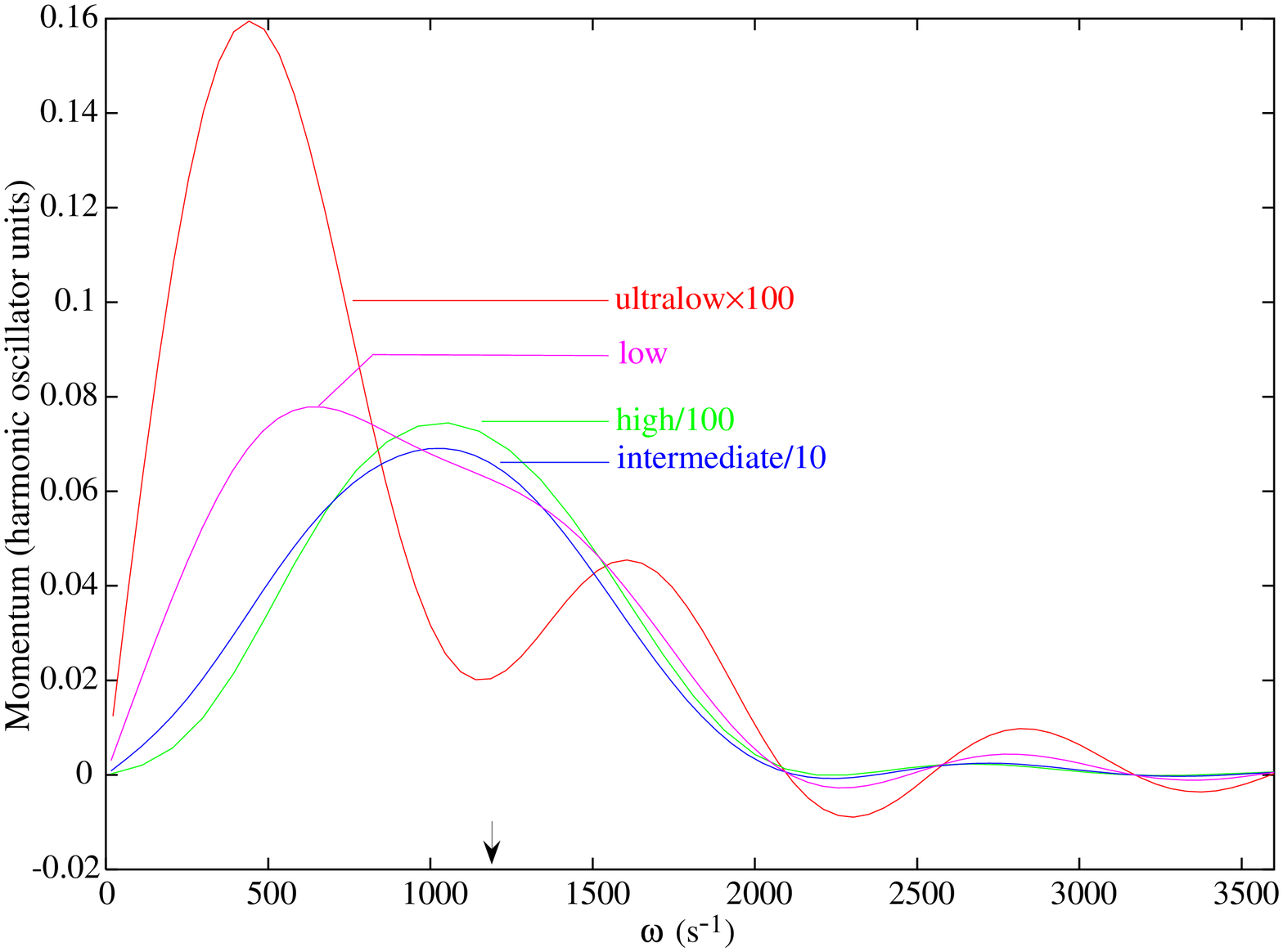}}
\caption {Same as Fig.~\ref{fig1} except for 3$^{\text{o}}$ ($q = 5.24
\times 10^{-2} \, \hbar k_{ph}$).  Results for four different laser
intensities corresponding to potential well depths $V_0 = 0.00162$,
0.0054, 0.0162 and 0.054 $E_R$ are shown.  The arrow indicates position of
$\Omega_B(q)$.}
\label{fig8}
\end{figure}

Fig.~\ref{fig8} shows $P_z$ versus $\omega$ for $\theta =
3^{\text{o}}$ ($q = 5.24 \times 10^{-2} \, \hbar k_{ph} = 11
\sqrt{\hbar m \omega_z}$), $t_p = 6$ ms and $V_0 = 0.00162$, 0.0054,
0.0162 and 0.054 $E_R$.  The peak near $\omega = \Omega_B(q) = 1234$
s$^{-1}$ for intermediate and high intensities becomes broadened at
low intensity and then splits into two peaks at $\omega \approx 400$
and 1500 s$^{-1}$ for ultralow intensity, with the one at $\omega
\approx 400$ s$^{-1}$ is about four times larger in magnitude than the
one at $\omega \approx 1500$ s$^{-1}$.  The dip between the two peaks
for the ultralow intensity is apparently also an interference effect
that can not be explained in terms of a local density approximation to
the Bogoliubov spectrum.  For intermediate and high intensities, the
peaks in $|\psi(k_z)|^2$ at $k_z = q$ are sufficiently large that
interference with the structure around the $k_z = 0$ peak does not
occur and therefore the interference dip is absent.

\section{Summary and Conclusions}

For very low optical intensities, long pulse duration and sufficiently
large values of the momentum-transfer imparted by the light-potential,
the local density approximation for the Bogoliubov excitation spectrum
is a reasonable first approximation; the response peaks in
$P_z(q,\omega,V_0,t_p)$ versus $\omega$ are broadened, particularly to
lower values of $\omega$.  However, we have shown that even for
relatively low optical intensities, power broadening results and
higher order processes occur that correspond to moving the atoms from
the wave packet with central momentum $k_z = q$ to wavepackets with
central momentum $2q$ and $0$ as well as additional $k_z = nq$, with
$n > 2$ and $n \leq -1$.  Moreover, at lower values of
momentum-transfer $q$ (smaller angles $\theta$) where the wavelength
of the optical potential becomes comparable to or larger than the size
of the condensate, interference effects play a role in the dynamics,
and directly affect the spectrum, $P_z(q,\omega,V_0,t_p)$ versus
$\omega$ and can produce additional maxima and minima in the spectrum. 
The power spectrum of the order parameter after the optical potential
is dropped, $|\psi(k_z,t_p)|^2$ versus $k_z$, can be used to
understand the nature of the spectra $P_z(q,\omega,V_0,t_p)$ versus
$\omega$ and $q$.

One should keep in mind that the study performed here is a 1D
calculation of the full 3D dynamics; 3D effects may modify details of
the results we obtained.  Nevertheless, we believe that the
qualitative features of the conclusions will not change.  We have
detailed elsewhere how our quasi-1D calculations of the type we
presented here model 3D aspects of the dynamics in cylindrically
symmetric potentials \cite{BTM_02}, but this method can not describe
radial excitations of the BEC that might arise due to the optical
potential via the mean-field interaction.  To the extent that radial
excitations are not important, our method should be an adequate
approximation to the 3D dynamics.

\begin{acknowledgments} 
We are extremely grateful to the group of Nir Davidson at Weizmann for
sharing information about their experiments and for useful
discussions.  This work was supported in part by grants from the
U.S.-Israel Binational Science Foundation (grant No.~98-421),
Jerusalem, Israel, the Israel Science Foundation (grant No.~212/01)
and the Israel MOD Research and Technology Unit.
\end{acknowledgments}

% \begin{figure}[!htb]
% \centerline{\includegraphics[width=3in,angle=270,keepaspectratio]{N_f7.eps}}
% \caption {$|\psi(k_z,t)|^2$ vs.~$k_z$, for the ultralow intensity case
% shown in Fig.~\ref{fig6} at three values of $\omega = 370$, 990 and
% 1687 s$^{-1}$.  The units of $k_z$ are in harmonic oscillator momentum
% units, $\sqrt{\hbar m \omega_z} = 2.44 \times 10^{-3} \, \hbar
% k_{ph}$.}
% \label{fig7}
% \end{figure}
% 

\end{document}